\documentclass[floatfix,twocolumn,showpacs,aps,tightenlines]{revtex4-1}
\usepackage{graphicx}
\usepackage{color}
\usepackage{psfrag}
\usepackage{dcolumn}% Align table columns on decimal point
\usepackage{bm}% bold math

\usepackage{amssymb}
\usepackage{amsmath}
\usepackage{amsfonts}
\usepackage{longtable}
\usepackage{xspace}
\usepackage{mathrsfs}
\usepackage[export]{adjustbox}

\newcommand{\beq}{\begin{equation}}
\newcommand{\eeq}{\end{equation}}

\newcommand{\be}{\begin{equation}}
\newcommand{\ee}{\end{equation}}
\newcommand{\bea}{\begin{eqnarray}}
\newcommand{\eea}{\end{eqnarray}}
\newcommand{\bes}{\begin{subequations}}
\newcommand{\ees}{\end{subequations}}

\newcommand{\scri}{\mathscr{I}}

\usepackage{bm}

\begin{document}

\title{The Third RIT binary black hole simulations catalog}
\author{James Healy}
\author{Carlos O. Lousto}
\affiliation{Center for Computational Relativity and Gravitation,
School of Mathematical Sciences,
Rochester Institute of Technology, 85 Lomb Memorial Drive, Rochester,
New York 14623}

\date{\today}

\begin{abstract}
The third release of the RIT public
catalog of numerical relativity black-hole-binary
waveforms \url{http://ccrg.rit.edu/~RITCatalog}
consists of 777 accurate simulations that include 300 precessing
and 477 nonprecessing binary systems with mass ratios $q=m_1/m_2$ in the
range $1/15\leq q\leq1$ and individual spins up to $s/m^2=0.95$.
The catalog also provides
initial parameters of the binary, 
trajectory information, peak radiation, and final remnant black hole
properties. 
The waveforms are corrected for the center of mass drifting
and are extrapolated to future null infinity. We successfully 
test this correction comparing with simulations of low radition 
content initial data.
As an initial application of this waveform catalog we reanalyze all the
peak radiation and remnant properties to find new, simple,
correlations among them for practical astrophysical usage.
%Direct use for parameter estimation of gravitational wave signals 
%is given in a companion paper.
\end{abstract}

 \maketitle

\section{Introduction}\label{sec:Intro}

Since the 
breakthroughs~\cite{Pretorius:2005gq,Campanelli:2005dd,Baker:2005vv} in
numerical relativity that solved the binary black hole problem
those techniques have been used to explore the late
dynamics of spinning black-hole binaries beyond the post-Newtonian
regime for several years.
The first generic, long-term precessing
black-hole binary evolutions (i.e., without any symmetry) were
performed in Ref.~\cite{Campanelli:2008nk}, where a detailed
comparison with post-Newtonian $\ell=2,3$ waveforms was made.
Numerical simulations have then started to explore the corners of parameter space,
these include near extremal \footnote{$\chi$ denotes the spin angular
momentum of a black hole in units of the square of its mass.  The
maximum possible spin is $\chi=1$} with $\chi=0.99$ spinning black-hole
binaries in Refs.~\cite{Lovelace:2014twa,Zlochower:2017bbg},
mass ratios as small as 
%$q=1/100$ in Refs.~\cite{Lousto:2010ut,Sperhake:2011ik},
$q=1/128$ in Ref.~\cite{Lousto:2020tnb},
and large initial separations, $R=100m$, in Ref.~\cite{Lousto:2013oza},
as well as very long waveforms starting at proper separations of $25m$
for a precessing binary in \cite{Lousto:2015uwa}
and for a nonspinning binary with 176 orbits in Ref.~\cite{Szilagyi:2015rwa}.

Other important studies include the exploration of the {\it hangup}
effect, i.e. the role individual black-hole spins play to delay or
accelerate their merger \cite{Campanelli:2006uy, Hannam:2007wf,Hemberger:2013hsa,Healy:2018swt}, the determination of the magnitude
and direction of the {\it recoil} velocity of the final merged black
hole \cite{Campanelli:2007ew,Campanelli:2007cga,Herrmann:2007ex,Pollney:2007ss,Baker:2006vn,Gonzalez:2007hi, Schnittman:2007ij,Lousto:2011kp},
and the {\it flip-flop} of individual spins during the orbital phase
\cite{Lousto:2014ida, Lousto:2015uwa, Lousto:2016nlp}, as well as
precession dynamics \cite{Schmidt:2010it,Lousto:2013vpa,Pekowsky:2013ska,Ossokine:2015vda,Lousto:2018dgd} and the inclusion
of those dynamics to construct surrogate models for gravitational
waveforms \cite{Blackman:2017dfb,Blackman:2017pcm,Varma:2018mmi}.

Numerical relativity breakthroughs~\cite{Pretorius:2005gq,Campanelli:2005dd,Baker:2005vv}
led to detailed predictions of the gravitational waves from the late inspiral,
plunge, merger, and ringdown of black-hole-binary systems (BHB).
These predictions helped to accurately identify the first direct
detection \cite{TheLIGOScientific:2016wfe} of gravitational waves with
such binary black hole systems \cite{Abbott:2016blz,Abbott:2016nmj,TheLIGOScientific:2016pea,Abbott:2016wiq} and match them to 
targeted supercomputer simulations \cite{Abbott:2016apu,TheLIGOScientific:2016uux,Lovelace:2016uwp}.

There have been several significant efforts to coordinate numerical
relativity simulations to support gravitational wave observations.
These include the numerical injection analysis (NINJA) project
\cite{Aylott:2009ya, Aylott:2009tn, Ajith:2012az, Aasi:2014tra}, the
numerical relativity and analytical relativity (NRAR) collaboration
\cite{Hinder:2013oqa}, and the waveform catalogs released by the
SXS collaboration~\cite{Mroue:2013xna,Blackman:2015pia, Chu:2015kft,Boyle:2019kee},
Georgia Tech.~\cite{Jani:2016wkt}, and RIT~\cite{Healy:2017psd,Healy:2019jyf}.

In this paper we describe a new release of the public waveform catalog
by the RIT numerical relativity group that total 777
simulations by adding a new set of 457 waveforms,
with 203 aligned spins and 254 precessing binaries.
The catalog includes all waveform modes $\ell\leq4 $ modes of $\psi_4$ and the strain $h$
(both extrapolated to null-infinity) and is updated to correct for
the center of mass displacement during inspiral and after merger.
The catalog can be accessed from \url{http://ccrg.rit.edu/~RITCatalog}.

This paper is organized as follows.  In Section \ref{sec:FN} we
briefly describe the methods and criteria for producing the numerical
simulations, the new center of mass correction and evaluation of
their errors in order to be included in
the RIT catalog.  In Sec.~\ref{sec:Catalog}
we describe the relevant BHB parameters, the file format, and the
content of the data in the catalog. A more detailed set of the results
is given in Sec.~\ref{sec:correlations} and in the appendix. We conclude in
Sec.~\ref{sec:Discussion} with a discussion of the future use of this
catalog for parameter inference of new gravitational waves events and
the extensions to this work to longer, more generic precessing binaries.

\section{Full Numerical Evolutions}\label{sec:FN}

The simulations in the RIT Catalog were evolved using the {\sc
LazEv} code~\cite{Zlochower:2005bj} implementation of the moving puncture
approach~\cite{Campanelli:2005dd} (with the modifications suggested by
Ref.~\cite{Marronetti:2007wz}). In all cases (except the very high spin
where we use CCZ4~\cite{Alic:2011gg}) we use the BSSNOK
(Baumgarte-Shapiro-Shibata-Nakamura-Oohara-Kojima) family
of evolutions systems~\cite{Nakamura87, Shibata95, Baumgarte99}.
For the runs in the catalog, we used a variety of finite-difference
orders, Kreiss-Oliger dissipation orders, and Courant factors~\cite{Lousto:2007rj,Zlochower:2012fk,Healy:2016lce}.
All of these are
given in the metadata included in the catalog and the references associated
with each run (where detailed studies have been performed). 

The {\sc LazEv} code uses the {\sc Cactus}~\cite{cactus_web} / {\sc Carpet}~\cite{Schnetter-etal-03b} / 
{\sc EinsteinToolkit}~\cite{Loffler:2011ay, einsteintoolkit} 
infrastructure.  The {\sc Carpet} mesh refinement driver provides a
``moving boxes'' style of mesh refinement. In this approach, refined
grids of fixed size are arranged about the coordinate centers of both
holes.  The code then moves these fine grids about the computational
domain by following the trajectories of the two black holes (BHs).

We use {\sc AHFinderDirect}~\cite{Thornburg2003:AH-finding} to locate
apparent horizons.  We first measure the magnitude of the horizon spin 
using the {\it isolated horizon} (IH) algorithm detailed in
Ref.~\cite{Dreyer02a} (as  implemented in
Ref.~\cite{Campanelli:2006fy}).
Once we have the horizon spin, we can calculate the horizon
mass via the Christodoulou formula 
${m_H} = \sqrt{m_{\rm irr}^2 + S_H^2/(4 m_{\rm irr}^2)}\,,$
where $m_{\rm irr} = \sqrt{A/(16 \pi)}$ and $A$ is the surface area
of the horizon. 

To compute the numerical (Bowen-York) initial data, we use the puncture
approach~\cite{Brandt97b} along with the {\sc
  TwoPunctures}~\cite{Ansorg:2004ds} code.  To compute initial low
eccentricity orbital parameters, we use the 3rd. post-Newtonian techniques
described in~\cite{Healy:2017zqj} to determine quasi-circular orbits.
We then evaluate the residual
eccentricity during evolution via the simple formula,
as a function of the separation of the holes, $d$,
$e_d=d^2\ddot{d}/m$, as given in \cite{Campanelli:2008nk}.

To generate more realistic initial data with reduced spurious 
gravitational wave content, in Ref.~\cite{Ruchlin:2014zva} we have chosen
a background ansatz as a conformal superposition of (possibly boosted)
Kerr spatial metrics. 
These new initial data, labeled as HiSpID when compared with the well 
known Bowen-York solutions, produce up to an order of magnitude 
reduction in the initial unphysical gravitational radiation signature. 
Those HiSpID data are relevant for nonspinning as well as very 
highly spinning black holes in a binary~\cite{Zlochower:2017bbg}, 
and high energy 
collisions~\cite{Healy:2015mla}. To generate those data we generalize the 
{\sc TwoPunctures} code~\cite{Ansorg:2004ds} to solve a coupled system of the Hamiltonian and
momentum constraints. We can thus evolve arbitrarily highly spinning 
black holes in quasi-circular orbits with unequal masses and different 
spin orientations~\cite{Healy:2017vuz}. We will use these data for evolving
highly spinning binaries with intrinsic spins $\alpha_i=S_i/m_i^2>0.9$.

As discussed in Ref. \cite{Healy:2017psd} the main sources
of numerical errors in this catalog are 
due to finite difference truncation, finite extraction
radii, use of finite number of modes, 
the non-zero residual initial eccentricities, and displacement
of the center of mass.
During the early inspiral, the irreducible masses and intrinsic spins
of each black hole should be nearly constant because the levels of
gravitational
wave energy and momentum absorbed by the holes is 4-5 orders of
magnitude smaller\cite{Isoyama:2017tbp} than those emitted to
infinity. During a simulation, the masses and spins variations are dominated by
numerical truncation error. We then use these variations on the horizon
masses and spins as a measure of the size of the truncation error.
For our current simulations we monitor accuracy by measuring
the conservation of the individual horizon masses and spins during evolution,
as well as the level of satisfaction of the Hamiltonian and 
momentum constraints,
to ensure reaching an accuracy consistent with our main applications.
Those measurements are seen to be preserved during evolution to at least one 
part in $10^{4}$ in the cases of the masses and one part in $10^3$ in the 
cases of the spins (see for instance Fig. 6 in Ref.~\cite{Lousto:2018dgd}). 

We measure radiated energy,
linear momentum, and angular momentum, in terms of the radiative Weyl
Scalar $\psi_4$, using the formulas provided in
Refs.~\cite{Campanelli:1998jv, Lousto:2007mh}. 
These formulas are
strictly speaking only valid at future null-infinity ($\scri^+$).
We therefore measure the radiated energy-momentum on a series of
timelike worldtubes  of finite
 radius and then extrapolate to $r=\infty$ using both linear and
quadratic extrapolations. The difference between these two
extrapolations give us an estimate for the uncertainty.

Unlike the radiated energy-momentum, more care is needed to properly
extrapolate the waveform itself to $\scri^+$.
As described in
Ref.~\cite{Nakano:2015pta},
we use the Teukolsky equation to obtain expressions for $r \psi_4$ at
$\scri^+$
based on its values on a timelike worldtube traced out by a fixed sphere of
constant (large) areal radius $r$ [see Eq. (29), there].
The expressions there contain the corrections of order ${\cal O}(1/r)$
and ${\cal O}(1/r^2)$ to $r \psi_4$. As shown in
Ref.~\cite{Nakano:2015pta}, this extrapolation is consistent with both
the waveform and the radiated energy-momentum extrapolated
using a least squares fit to a polynomial in $1/r$ powers.
Additionally, the ${\cal O}(1/r)$
perturbative corrections were shown to be consistent with a
Cauchy-Characteristic extraction for an equal-mass binary
in~\cite{Babiuc:2010ze}.

Various simulations in this catalog were studied in
detail in previous papers. 
In Appendix A of Ref.~\cite{Healy:2014yta}, we
performed a detailed error analysis of configurations with
equal mass and spins aligned/antialigned with respect to the orbital
angular momentum; 
%We varied the initial separation of the binary, the
%resolutions, grid structure, waveform extraction radii, and the number
%of $\ell$ modes used in the construction of the radiative quantities.
in Appendix B of Ref.~\cite{Healy:2016lce}, we performed
convergence studies for runs with mass ratios $(q=1,3/4,1/2,1/3)$ and
measured errors due to finite observer locations; 
and in Ref.~\cite{Healy:2017mvh}, we performed convergence studies
for $q\geq 1/10$ nonspinning binaries. In Ref.~\cite{Healy:2020iuc}
we make detailed convergence and gauge studies on nonspinning
$(q=1/2,1/3,1/5,1/3)$ and spinning $q=1$ binaries.

Finally, in addition to all the internal consistency analysis and
error estimates, in Ref.~\cite{Lovelace:2016uwp} we showed that for
the parameter estimated for GW150914 ($q=m_1/m_2=0.82$ and spins for
the small/large holes of $\chi_1=-0.44$ and $\chi_2=+0.33$), the RIT
waveforms and those produced completely independently by the SXS
collaboration have an excellent match~\cite{Cho:2012ed} of $\gtrsim 0.99$ overall
for modes up to $\ell=5$. In
Ref.~\cite{Healy:2017abq} a similar agreement between approaches has
been found for other five targeted precessing and nonprecessing simulations
of GW170104, displaying a 4th order convergence with finite difference
resolution. The comparisons were also carried up to $\ell=5$-modes.
For all modes up to $l\leq4$ we found a match of $\gtrsim 0.99$ 
and $\gtrsim 0.97$ for the $l=5$ modes.

In all our studies we concluded that the waveforms at the resolutions
provided in this catalog are well into the convergence regime (roughly
converging at 4th-order with resolution), that the horizon evaluated
quantities such as the remnant final mass and spins have errors of the
order of 0.1\%, and that the radiatively computed quantities such as
the recoil velocities and peak luminosities are evaluated at a typical
error of 5\%.

\subsection{Center of mass displacement}\label{sec:CofM}

Since full numerical simulations of binary black hole are started at
a finite separation they include non-physical initial data effects,
such as radiation content absorved by the holes adding to their 
masses and spins (see Tables \ref{tab:ID}, and \ref{tab:IDr} 
and \ref{tab:IDr_prec}, for initial versus settled values), and
an initial kick due to the net linear momuentum of the initial 
radiation content. Corrections by
this effect are normally performed in the simulation coordinates, hence
depend on the gauge chosen. The choice itself of the center of mass
and its gauge dependence have been studied in full general relativity
(See for instance the recent paper \cite{Kozameh:2019bnx} 
and references there in) 
and in the post-Newtonian approximation 
(See for instance the recent paper \cite{Kozameh:2017qiw,Bernard:2017ktp} 
and references there in).
In \cite{Woodford:2019tlo} a detailed characterization and study 
has been performed to 
evaluate the impact of a shifting center or mass during full numerical
evolutions has on the mode decomposition of the gravitational waveforms,
even if extrapolated to infinite resolution and infinite observer location.

\begin{figure}[h!]
  \includegraphics[angle=0,width=0.85\columnwidth]{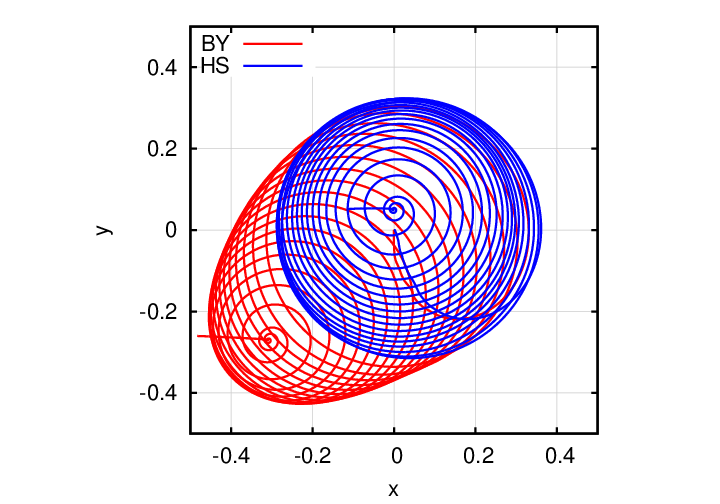}
  \caption{Wobbling and displacement of the center of mass for a binary with 
mass ratio $q=0.4142$ and spins $a_1=-0.50$ and $a_2=0.85$ for Bowen-York (BY) and HiSpID (HS) initial data.
      \label{fig:CofMq04142}}
\end{figure}

The center of mass shows two notable features when displayed in the simulations
coordinates: A wobbling of increasing radius due to the physical radiation 
of momenta during the inspiral motion and a displacement of the center of this 
wobbling away from the origin of coordinates (chosen formally at the initial 
center of mass location). 
See Figs.~\ref{fig:CofMq04142} and \ref{fig:CofMq08500}, and
note that both trajectories begin at the origin of coordinates $(x,y)=(0,0)$.
A practical implementation of the center of mass correction is to account
for a linear shift in time of the origin of coordinates for the multipole
decomposition of the waveform. This shift can be performed a posteriori,
by de-mixing modes at each time step from a center of coordinates that is 
linearly moving, interpolating between the initial configuration and the
location of the merger in coordinates. This linear correction is being applied
as an extrapolation post merger, since the post merger final black hole
acquires a physical recoil velocity due to a net radiation of linear momentum
carried in the form of gravitational waves to infinity. %\cite{}.
%\NOTE{Aren't we overdoing it? It seems to work comparing BY and HS!}

We have applied this correction to all the 777 
waveforms released in this catalog.
In order to exemplify its effectiveness, we have considered the evolution
of binary black holes with the usual Bowen-York initial data
\cite{Bowen:1980yu} and
compared the corresponding evolution of the HiSpID data 
\cite{Ruchlin:2014zva} with
much lower radiation content. We then applied the above correction to
the former simulation to see if this brings the mode contents toward those
of the later simulation as a test of effectiveness of the shift of the
center of mass correction.

\begin{figure}[h!]
  \includegraphics[angle=0,width=0.494\columnwidth]{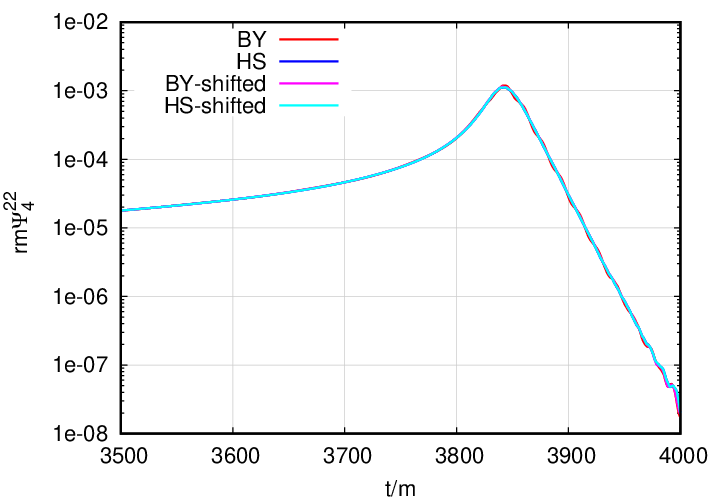}
  \includegraphics[angle=0,width=0.494\columnwidth]{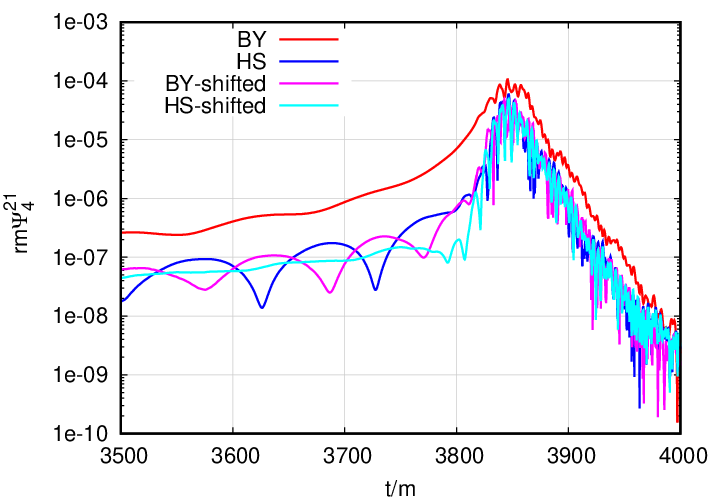}\\
    \includegraphics[angle=0,width=0.494\columnwidth]{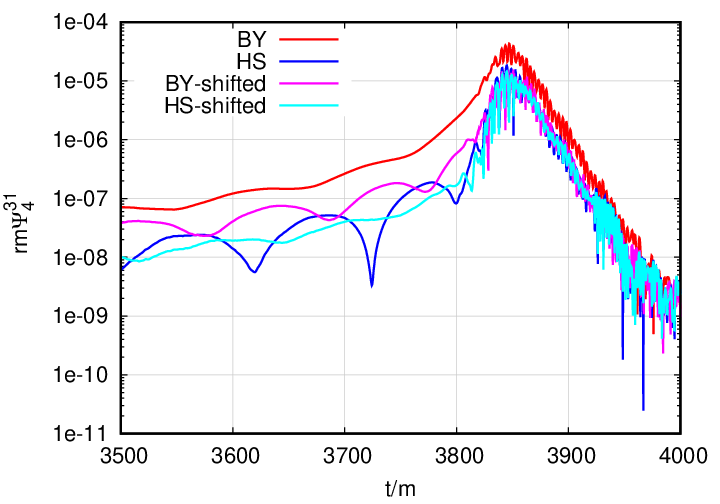}
    \includegraphics[angle=0,width=0.494\columnwidth]{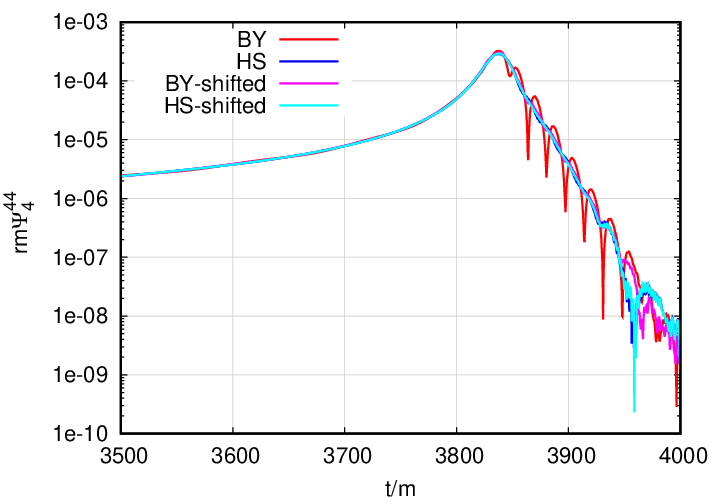}
  \caption{Amplitude of the modes $(\ell,m)=(2,2),(2,1),(3,1),(4,4)$ of the 
Weyl scalar $\psi_4$ decomposition at the extraction radius $R=75m$. Displayed are the evolution of BY and HS data amplitude, and the correction to the center of mass shift to the BY and HS data evolutions. 
    \label{fig:CofMq04142Modes}}
\end{figure}

\begin{figure}[h!]
  \includegraphics[angle=0,width=0.85\columnwidth]{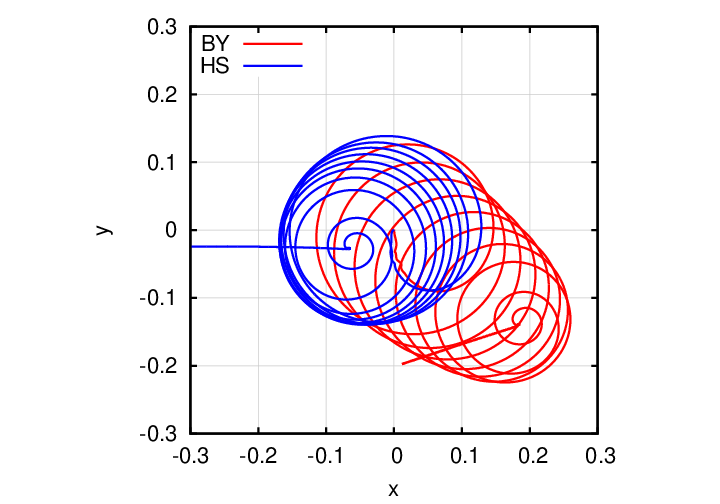}
  \caption{Wobbling and displacement of the center of mass for a binary with 
mass ratio $q=0.85$ and spins $a_1=-0.50$ and $a_2=-0.85$ for Bowen-York (BY) and HiSpID (HS) initial data.
      \label{fig:CofMq08500}}
\end{figure}

In Fig.~\ref{fig:CofMq04142} we display the wobbling and displacement of the center of mass for a nonprecessing binary with 
mass ratio $q=0.4142$ and spins $a_1=-0.50$ and $a_2=0.85$ for evolutions from Bowen-York (BY) and HiSpID (HS) initial data. Those show the much larger displacement of the BY data (roughly an order of magnitude) than for the HS data.
Thus we expect the later showing a much reduced mode mixing effect and providing a reference and measure of the benefits of applying the center of mass
shift correction to the BY data. This is indeed clearly the fact when as
shown in Fig.~\ref{fig:CofMq04142Modes} where we display a few selected modes of our 
$\ell_{max}=6$ analysis. The corrections to the BY data show a rectified amplitude
to the most affected modes during the inspiral $(\ell,m)=(2,1),(3,1)$ and
post merger  $(\ell,m)=(4,4)$ while $(\ell,m)=(2,2)$ appears mostly unaffected
at larger scale. The rest of the modes, not displayed here, fall in one
of the above categories. A further correction of the center of mass shift
can then be applied to the HS thus achieving even higher accuracy.

%\begin{widetext}
%\end{widetext}

\begin{figure}[h!]
  \includegraphics[angle=0,width=0.494\columnwidth]{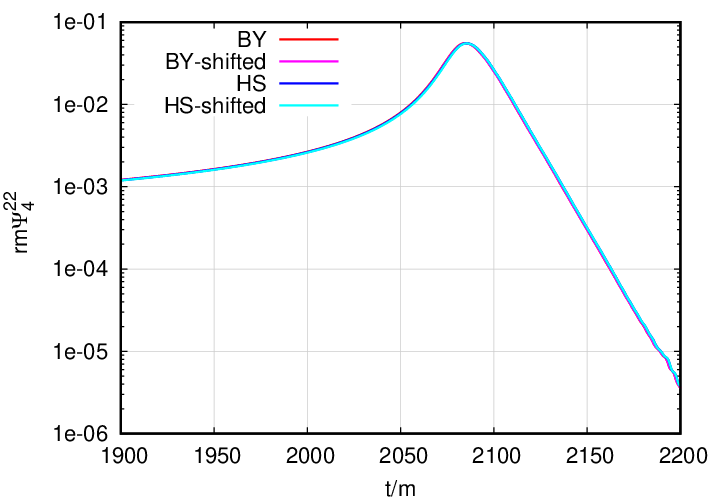}
  \includegraphics[angle=0,width=0.494\columnwidth]{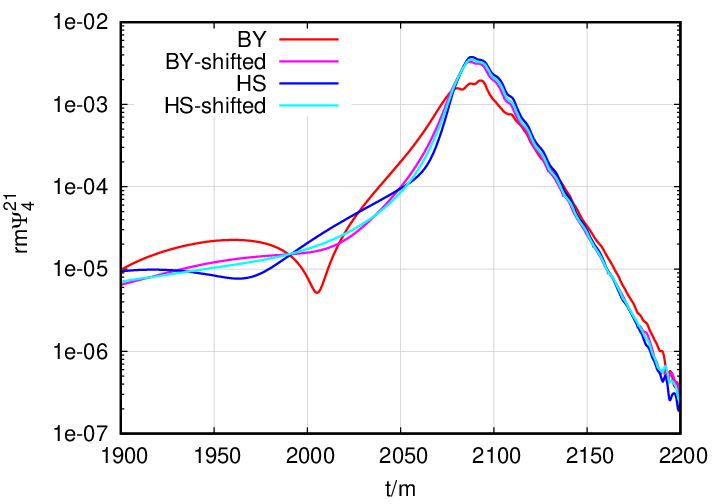}\\
    \includegraphics[angle=0,width=0.494\columnwidth]{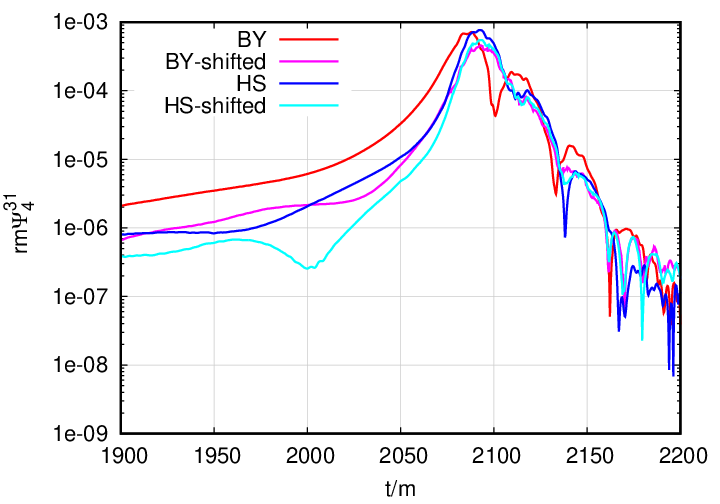}
    \includegraphics[angle=0,width=0.494\columnwidth]{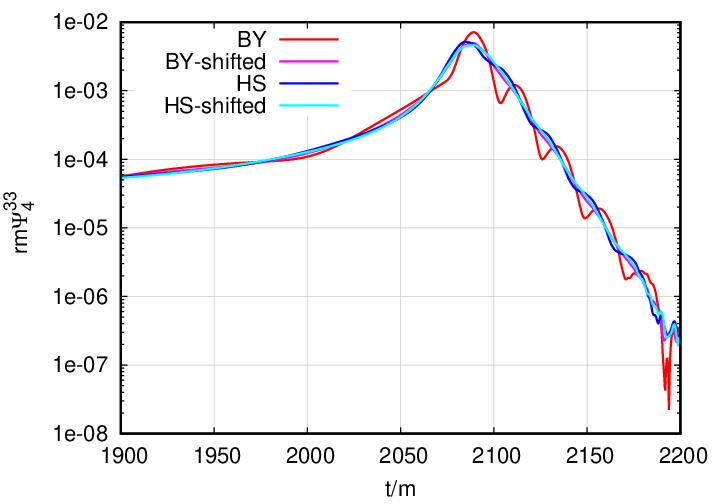}
  \caption{Amplitude of the modes $(\ell,m)=(2,2),(2,1),(3,1),(3,3)$ of the 
Weyl scalar $\psi_4$ decomposition at the extraction radius $R=113m$. Displayed are the evolution of BY and HS data amplitude, and the correction to the center of mass shift to the BY and HS data evolutions. 
    \label{fig:CofMq08500Modes}}
\end{figure}

In order to further cross check our results we have studied a second case
displayed in Fig.~\ref{fig:CofMq08500} showing the 
wobbling and displacement of the center of mass for a nonprecessing binary with 
mass ratio $q=0.85$ and spins $a_1=-0.50$ and $a_2=-0.85$ for evolutions from Bowen-York (BY) and HiSpID (HS) initial data. Those show the larger displacement of the BY data (roughly a factor three) than for the HS data.
Fig.~\ref{fig:CofMq08500Modes} displays similar results as the first simulation.

%\begin{widetext}
%\end{widetext}

%%%%%%

\section{The Catalog}\label{sec:Catalog}

%\begin{widetext}
\begin{figure*}
  \includegraphics[angle=0,width=0.494\columnwidth]{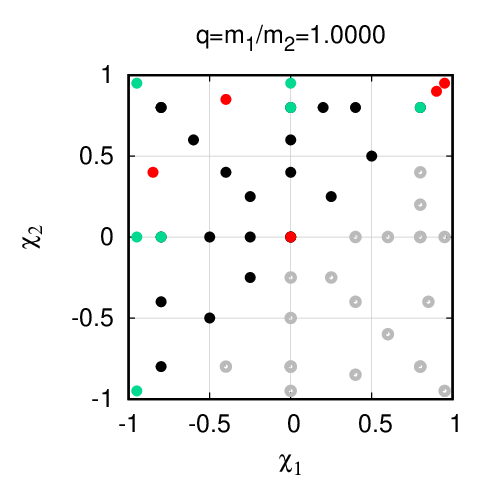}
  \includegraphics[angle=0,width=0.494\columnwidth]{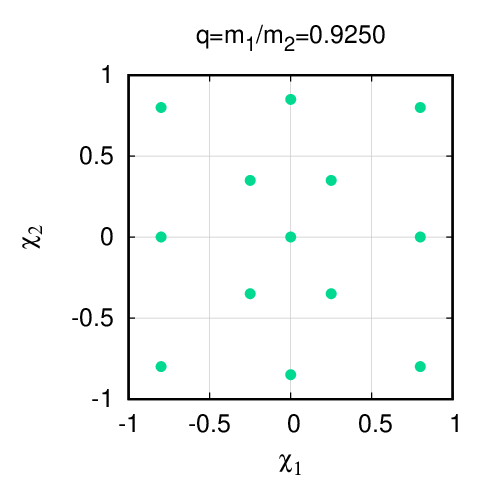}
    \includegraphics[angle=0,width=0.494\columnwidth]{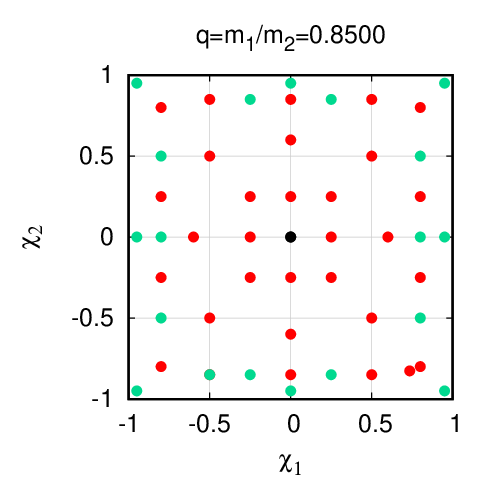}
    \includegraphics[angle=0,width=0.494\columnwidth]{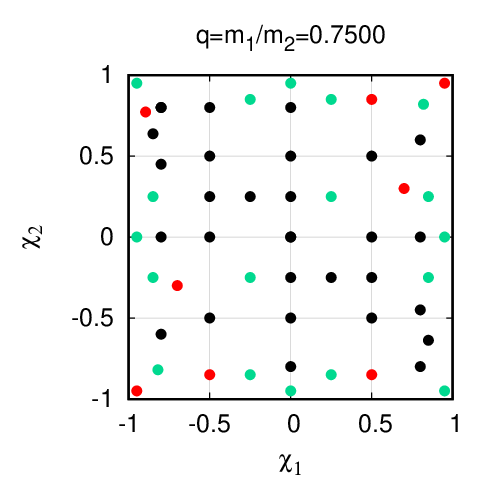}\\
    \includegraphics[angle=0,width=0.494\columnwidth]{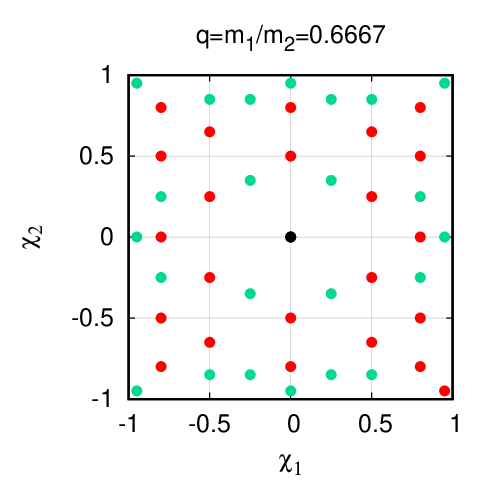}
    \includegraphics[angle=0,width=0.494\columnwidth]{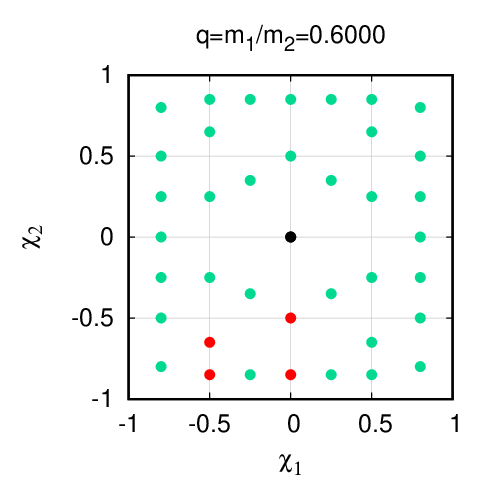}
    \includegraphics[angle=0,width=0.494\columnwidth]{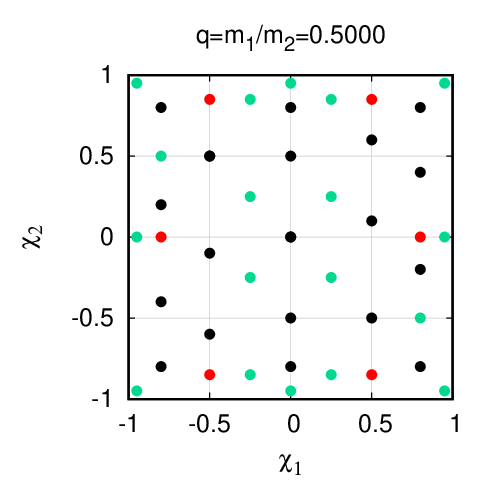}
  \includegraphics[angle=0,width=0.494\columnwidth]{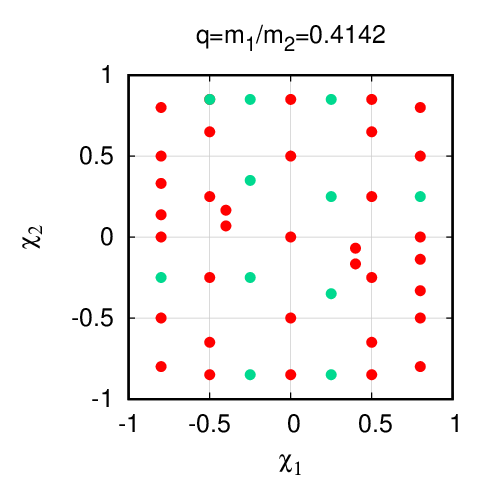}\\
    \includegraphics[angle=0,width=0.494\columnwidth]{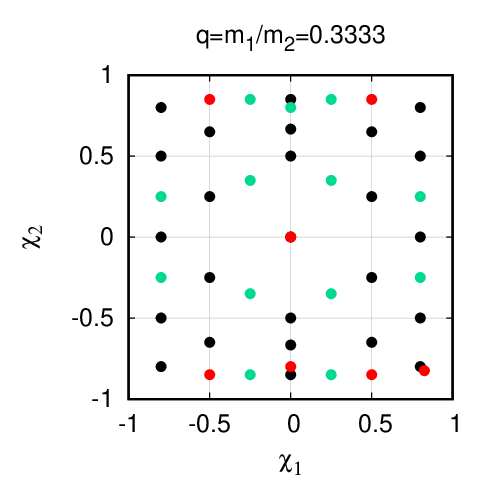}
    \includegraphics[angle=0,width=0.494\columnwidth]{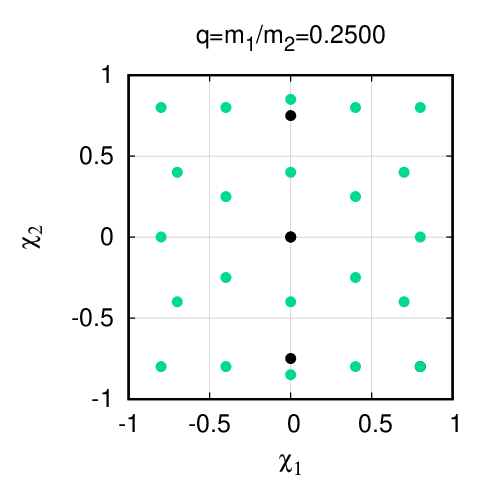}
      \includegraphics[angle=0,width=0.494\columnwidth]{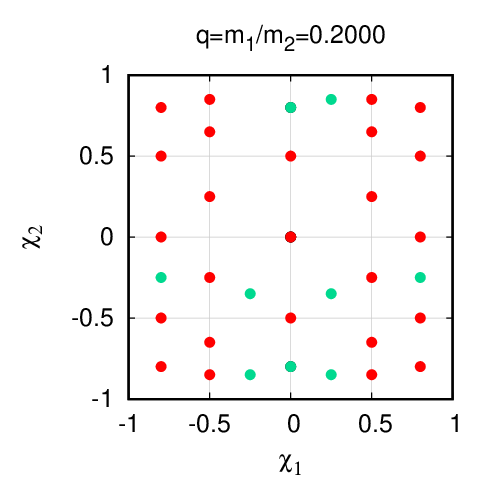}
  \caption{Initial parameters in the $(q,\chi_1,\chi_2)$ space
    for the 477 nonprecessing binaries. Note that $\chi_i$ denotes 
    the component of the dimensionless spin of BH $i$ along the
    orbital angular momentum.
    Each panel corresponds to a given mass ratio that covers the 
    comparable masses binary range from $q=1$ to $q=1/5$.
    The dots in black denote the simulations of the catalog first
    release, the dots in red are those of the second release,
    and the dots in green are those of this third release.
    \label{fig:panels}}
\end{figure*}
%\end{widetext}

%\begin{widetext}
\begin{figure}[h!]
  \includegraphics[angle=0,width=1.0\columnwidth]{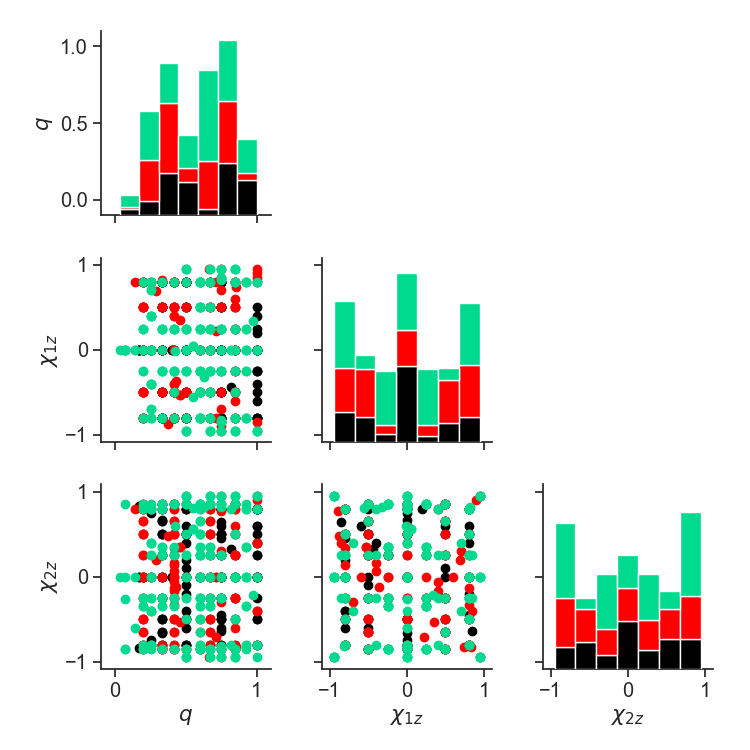}
  \caption{Counting simulations in the $(q,\chi_1,\chi_2)$ planes 
    (faces of the cube) for the 477 nonprecessing binaries.  The 120 release 1
    simulations are black, the 154 release 2 simulations are red and the 203 release 3 are in green.
      \label{fig:AlignedMulti}}
\end{figure}
%\end{widetext}

\begin{figure}[h!]
\includegraphics[width=0.7\columnwidth]{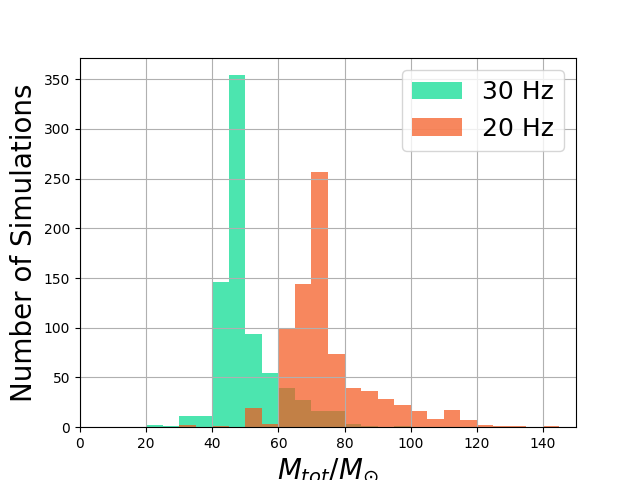}\\
  \includegraphics[width=0.7\columnwidth]{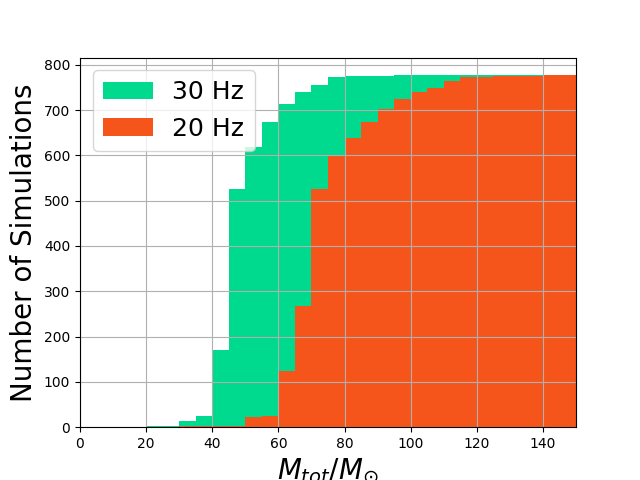}
  \caption{Top: Distributions of the total mass of BHB systems in the
RIT catalog corresponding to a starting gravitational wave frequency of
20 Hz (green) and 30 Hz (red) in bins of $5M_\odot$. 
Bottom: The cumulative version of the above plot also in bins of
$5M_\odot$ for the 777 simulations in this catalog.
\label{fig:mtotal}}
\end{figure}

The RIT Catalog can be found at \url{http://ccrg.rit.edu/~RITCatalog}.
Figure~\ref{fig:panels} shows the distribution of the
non-precessing runs in the catalog in
terms of $\chi_{1,2}$ and $q$ (where $\chi_i$ is the component of
the dimensionless spins of BH $i$ along the direction of the orbital angular
momentum).
The information currently in the catalog consists of the metadata
describing the runs and all modes up through the $\ell=4$ modes
(enough for most applications) of $m r \psi_4$ 
extrapolated to $\scri^+$ via the perturbative approach
of~\cite{Nakano:2015pta}. 
The associated metadata include the
initial orbital frequencies, ADM masses, initial waveform frequencies from (2,2) mode,  black hole masses, momenta, spins, separations,
and eccentricities, as well the black-hole masses and spins once the
initial burst of radiation has left the region around the binary.
Note that we normalize our data such that
the sum of the two initial horizon masses is $1m$.
These {\it relaxed} quantities (at $t_{relax}=200m$ after the initial burst
of radiation has mostly dissipated)
are more accurate and physically relevant for modeling purposes. 
In addition, we also include the final remnant black hole masses,
spins and recoil velocity.

The catalog is organized using an interactive table~\cite{datatables_web} that includes an
identification number, resolution, type of run (nonspinning, aligned spins,
precessing), the initial proper length of the coordinate 
line joining the two BH centroids that is outside both horizons~\cite{Lousto:2013oza},
the coordinate separation of the two centroids, the mass ratio of the two black
holes, the components of the dimensionless spins of the two black
holes, the starting waveform frequency, $m f_{22, {\rm relax}}$, 
time to merger, number of gravitational wave cycles 
calculated from the (2,2) modes from the beginning of the inspiral signal 
to the amplitude peak,
remnant mass, remnant spin, recoil
velocity, peak luminosity, amplitude and frequency.
The final column gives the appropriate
bibtex keys for the relevant publications where the waveforms were
first presented. The table can be sorted (ascending or descending) by
any of these columns. And there is a direct search feature that runs over all
table elements.

Resolutions are given in terms of the grid spacing of the refinement
level where the waveform is extracted (which is typically two
refinement levels below the coarsest grid) with $R_{obs}\sim100m$.
We use the notation 
nXYY, where the grid spacing in the wavezone is given by $h=m/X.YY$, e.g.,
n120 corresponds to $h=m/1.2$.

For each simulation in the catalog there are three files: one contains
the metadata information in ASCII format, the other two are a tar.gz files
containing ASCII files with up to and including $\ell=4$
modes of $m r\psi_4$ and $h$.  In the near future, data will be available in the
Numerical Relativity Injection format \cite{Schmidt:2017btt}. 
%\NOTE{decide what to include and update here.  Include extrap-com-shifted-psi4 tarball, extrap-com-shifted-cleaned-strain in LVCNR format?}
Note that the primary data in our catalog is the Weyl scalar $mr \psi_4$ extrapolated
to $\scri^+$  (using Eq.~(29) of Ref.~\cite{Nakano:2015pta}),
rather than the strain $(r/m)h$.
We provide the strain but also leave it to the user to
convert $mr\psi_4$ to strain for most modes
since this is still a sensitive process and is best handled on a
mode-by-mode basis.
The subtleties associated with transforming $\psi_4$ to $h$ arise from
the two integrations required \cite{Baker:2002qf,Campanelli:2008nk}.
One of the standard techniques, developed in Ref.~\cite{Reisswig:2010di},
performs this integration in Fourier space with a windowing function
and a low-frequency cutoff. Both of these require fine-tuning of
parameters. The codes to do this are open-source and publicly 
available from \url{https://svn.einsteintoolkit.org/pyGWAnalysis/trunk}.

Figure~\ref{fig:AlignedMulti} shows the distribution of the 477 non-precessing
runs in the catalog in terms of $\chi_{1,2}$ and $q$.
Those runs were  motivated by
systematic studies to produce a set of accurate remnant
formulas to represent the final mass, spin and recoil of a merged
binary black hole system and the peak luminosity, amplitude and
frequency, as a function of the parameters of the
precursor binary, as reported in 
\cite{Healy:2014yta,Healy:2016lce,Healy:2018swt}.
A second important motivation was to provide a grid of simulations for
parameter estimation of gravitational wave signals detected by LIGO
using the methods described in \cite{Abbott:2016apu}. 
%We will see in the next section that we have achieved a good coverage of this BHB parameter space.

\begin{figure*}
  \includegraphics[angle=0,width=0.65\columnwidth]{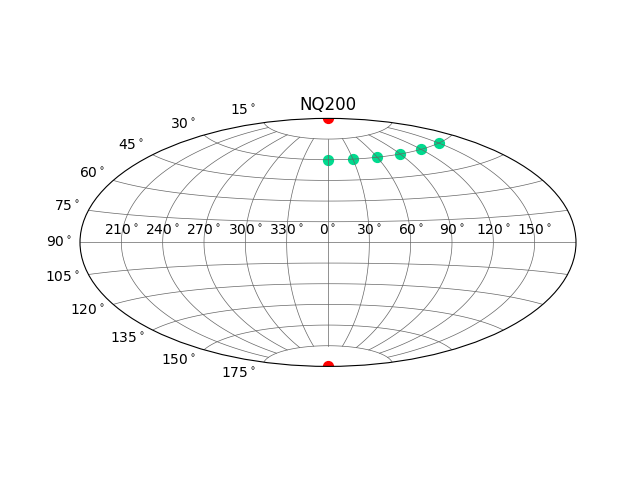}
  \includegraphics[angle=0,width=0.65\columnwidth]{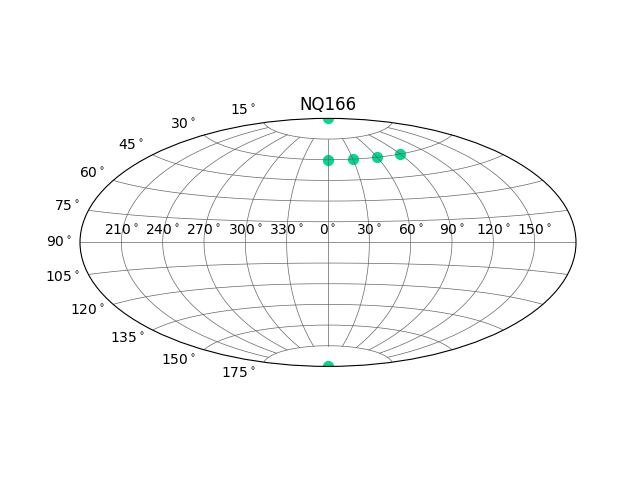}
    \includegraphics[angle=0,width=0.65\columnwidth]{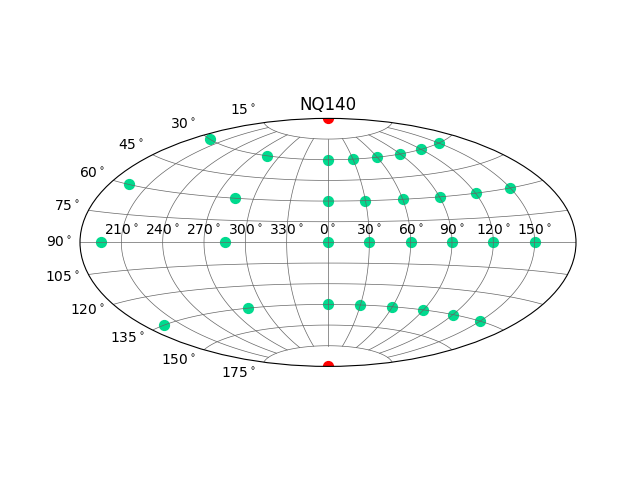}\\
    \includegraphics[angle=0,width=0.65\columnwidth]{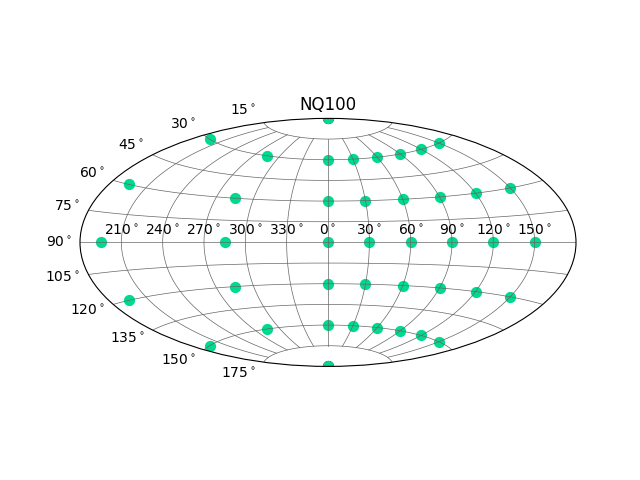}
    \includegraphics[angle=0,width=0.65\columnwidth]{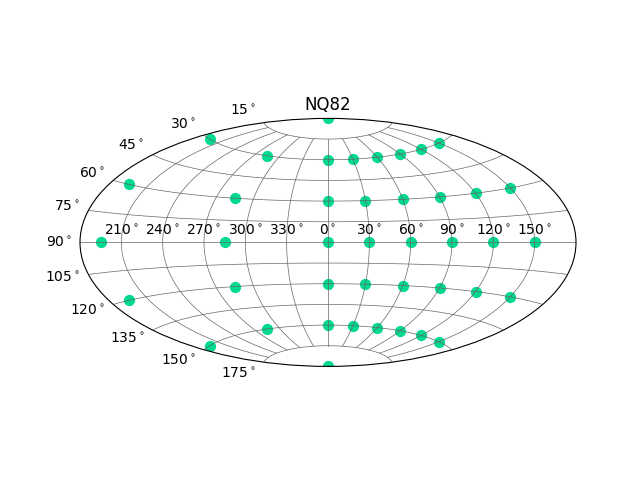}
    \includegraphics[angle=0,width=0.65\columnwidth]{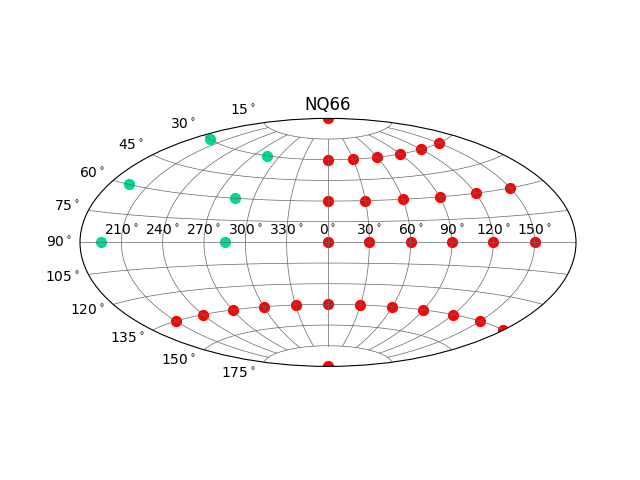}\\
    \includegraphics[angle=0,width=0.65\columnwidth]{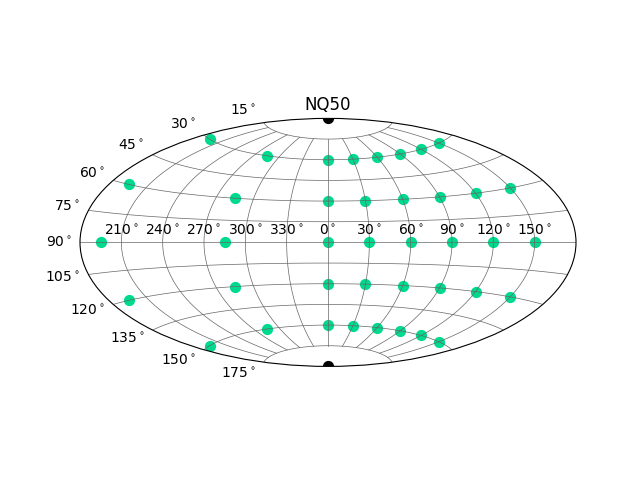}
  \includegraphics[angle=0,width=0.65\columnwidth]{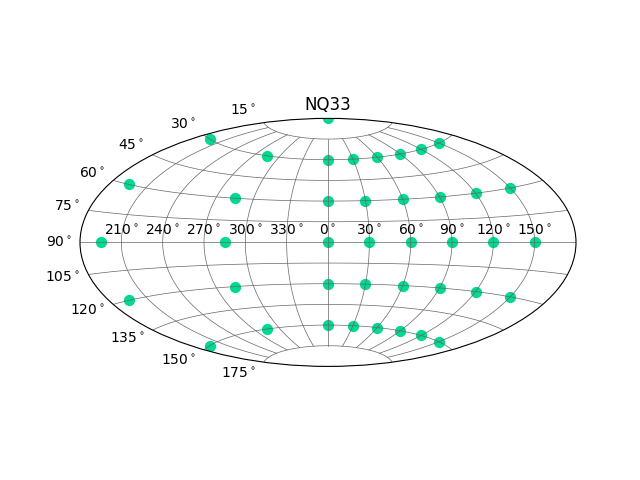}
    \includegraphics[angle=0,width=0.65\columnwidth]{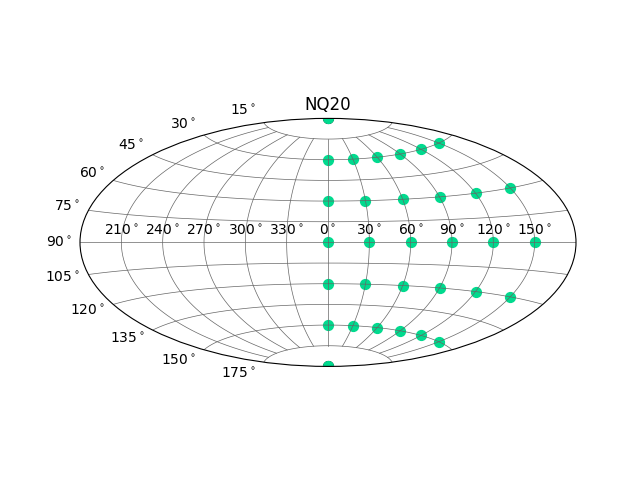}
  \caption{Initial parameters in the $(q,\theta_2,\phi_2)$ space
    for the precessing binaries. Note that $(\chi_2=0.8,\theta_2,\phi_2)$
    denotes 
    the component of the dimensionless spin of the BH $i=2$ from the direction
    of the orbital angular momentum.
    Each panel corresponds to a given mass ratio that covers the 
    comparable masses binary range $q=1, 0.82, 2/3, 1/3, 1/5$ and
    $q=2, 5/3, 1.4$, where $q>1$ means it is the smaller hole that is spinning.
    The dots in black denote the simulations of the catalog first
    release, the dots in red are those of the second release,
    and the dots in green are those of this third release.
    \label{fig:precpanels}}
\end{figure*}

The precessing runs in the catalog were motivated to study  
particular spin dynamics of merging BHB, such as
the study of unstable spin flip-flop,
as reported in \cite{Lousto:2016nlp} and the targeted followups
of gravitational wave signal from the first and second LIGO observing runs
\cite{Lovelace:2016uwp,Healy:2017abq}.
We have also paid special attention to the systematic study of simulations
covering a 4-dimensional parameter space involving a spinning and a nonspinning black hole binary as a function of the mass ratio (See Fig.~\ref{fig:precpanels}).
Those simulations were originally performed to study remnant recoil and final masses and spins \cite{Zlochower:2015wga}. We have supplemented them here with
additional simulations to have a coverage of spin orientations
(see Fig.~\ref{fig:precpanels} and Table \ref{tab:ID}) that allows an
estimation of precession).

%\begin{widetext}
%\end{widetext}

Figure~\ref{fig:mtotal} shows the distributions of the minimal total mass of the 
BHB systems in the catalog given a starting gravitational wave frequency
of 20 or 30 Hz in the source frame.
This provides a coverage for the current events observed by LIGO (redshift
effects improve this coverage by a factor of $1+z$, where $z$ is the
redshift). Coverage
of even lower total masses would require longer simulations or hybridization
of the current waveforms with Post-Newtonian methods~\cite{Ajith:2012az}.

\subsection{Non precessing merging binaries correlations}\label{sec:correlations}

Detailed and higher order formulas relating the binary parameters to
the post-merger properties of the final remnant black hole and merger
waveform have been studied in 
Refs. \cite{Healy:2014yta,Healy:2016lce,Healy:2018swt} using the
aligned spins simulations of previous releases of the RIT catalog.

\begin{figure*}
  \hfill\includegraphics[angle=0,width=0.65\columnwidth]{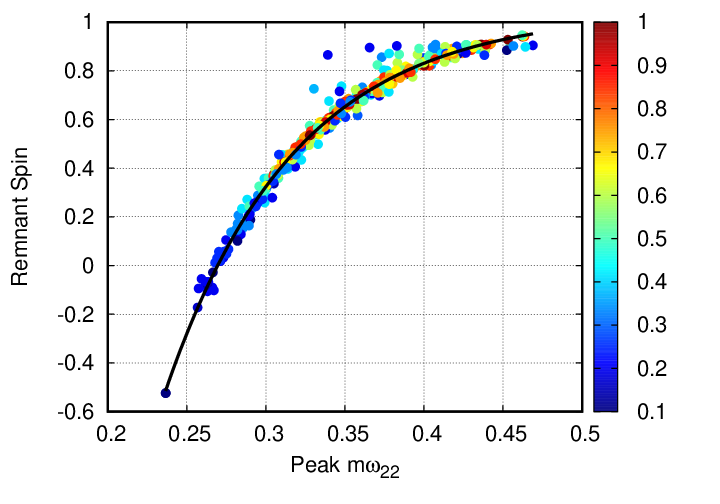}
  \includegraphics[angle=0,width=0.65\columnwidth]{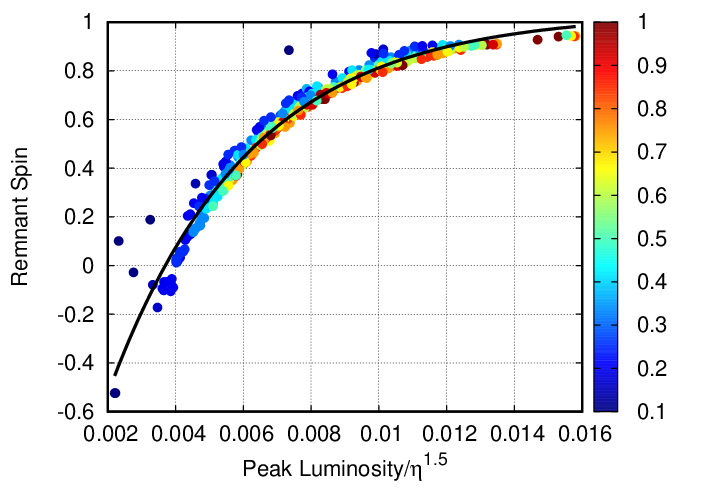}
  \includegraphics[angle=0,width=0.65\columnwidth]{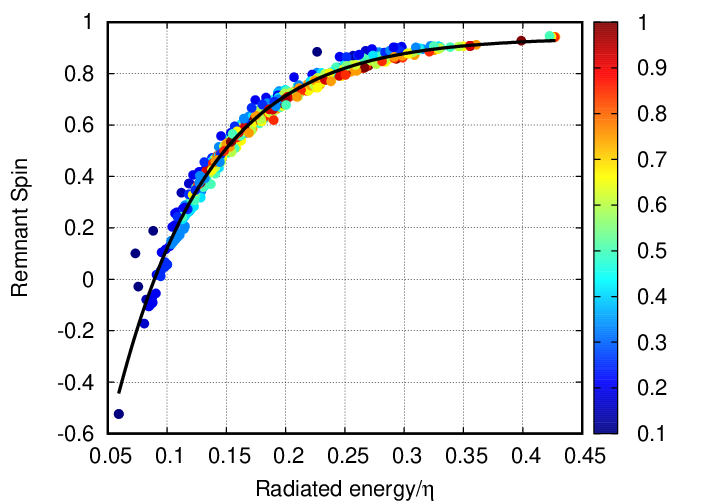}\\
  \hfill\includegraphics[angle=0,width=0.65\columnwidth]{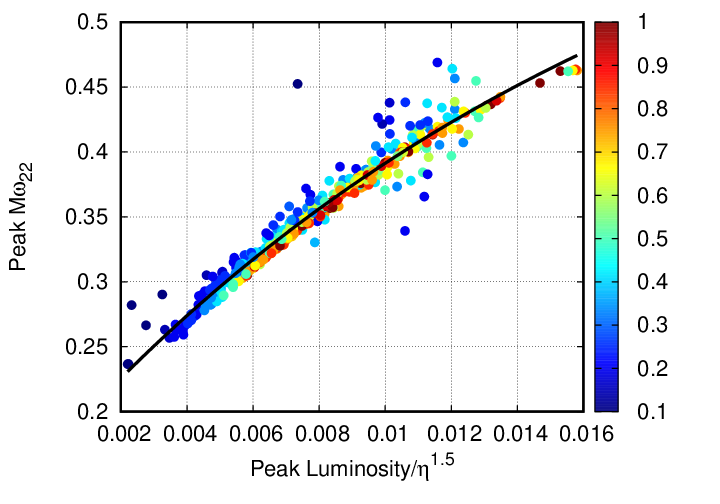}
  \includegraphics[angle=0,width=0.65\columnwidth]{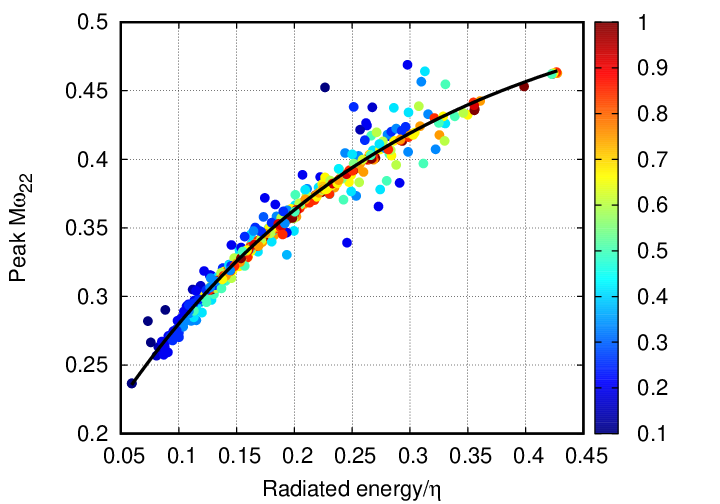}\\
   \hfill\includegraphics[angle=0,width=0.65\columnwidth]{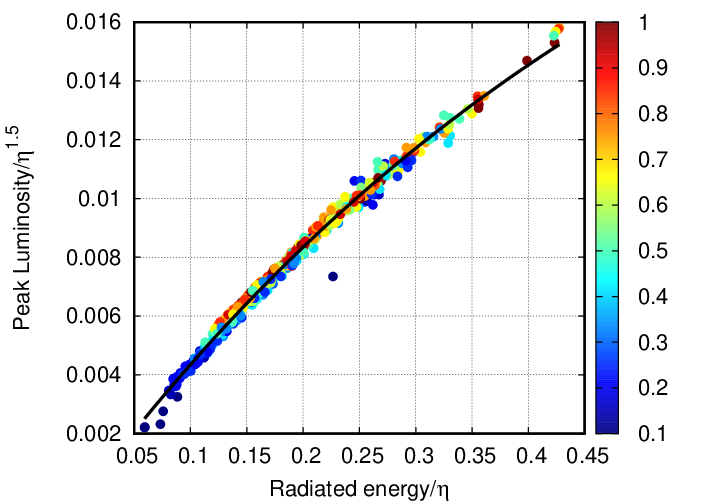}
  \caption{Correlations between the radiated energy, peak luminosity, peak
    frequency at merger and spin remnant with the corresponding scaling
    by the symmetric mass ratio to obtain the leading mass dependence.
          \label{fig:correlations}}
\end{figure*}

Foreseeing astrophysical applications and use in the future of massive catalogs
of binary black holes,
we display in Fig. \ref{fig:correlations}
simple scaling phenomenological
correlations between remnant parameters and merger waveforms.

%\begin{widetext}
%\end{widetext}

Fitting formulas for the simple correlations and estimated errors in
their coefficient are 
\bea\label{eq:correlations1}
%\begin{align}
  \alpha_f  = 1.044\pm0.008 - (27.76\pm1.37) \mathrm{e}^{(-12.17\pm0.20) m\omega_{22}}, \nonumber\\
  \alpha_f  = 1.037\pm0.011 - ( 2.54\pm0.04) \mathrm{e}^{(-242.69\pm5.90) \mathcal{L}/\eta^{1.5}}, \nonumber\\
  \alpha_f  = 0.940\pm0.005 - ( 2.96\pm0.05) \mathrm{e}^{(-12.84\pm0.19) E_{rad}/\eta},
%\end{align}
\eea
for the final spin $\alpha_f$ as a function of the peak frequency
$m\omega_{22}$ of the (2,2) mode, the peak Luminosity $\mathcal{L}$,
and the total radiated energy during merger $E_{rad}$
(normalized by the symmetric mass ratio $\eta=q/(1+q)$).

Similar correlations among those quantities can be found,
\bea\label{eq:correlations2}
%\begin{align}
m\omega_{22}  = 0.705\pm0.045 - (0.53\pm0.04) \mathrm{e}^{(-53.00\pm6.93) \mathcal{L}/\eta^{1.5}} \nonumber\\
m\omega_{22}  = 0.534\pm0.010 - (0.38\pm0.01) \mathrm{e}^{(-3.97\pm0.23) E_{rad}/\eta} \nonumber\\
\mathcal{L}/\eta^{1.5}  = 0.030\pm0.001 - (0.030\pm0.001)\mathrm{e}^{(-1.72\pm0.09) E_{rad}/\eta}.
%\end{align}
\eea

The fittings take the leading behavior and scaling of the correlations
without attempting higher order corrections to be used in astrophysical
estimates and control of more sophisticated implementations in large
catalogs of binary black hole gravitational wave signals and its modeling,
as well as tests of gravity.

%%%%%%%%%%%%%%

%%%% Carlos got up to here %%%%%%

%%%%%%%%%%%%%%%%%%%%%%%%%%%%%%%%%%%%%%%%

\section{Conclusions and Discussion}\label{sec:Discussion}

The breakthroughs~\cite{Pretorius:2005gq,Campanelli:2005dd,Baker:2005vv}
in numerical relativity were instrumental in identifying the first
detection of gravitational waves \cite{TheLIGOScientific:2016wfe} with
the merger of two black holes.  The comparison of
different approaches to solve the binary black hole problem has
produced an excellent agreement for the GW150914
\cite{Lovelace:2016uwp} and GW170104~\cite{Healy:2017abq}, including
higher (up to $\ell=5$) modes.
The use of numerical relativity waveform catalogs 
(See also Refs.~\cite{Abbott:2016apu,Lange:2017wki,Kumar:2018hml})
allows the application of a consistent
method for parameter estimation (of merging binary black holes) 
of the observed gravitational waves in the observational runs O1/O2.
The current aligned spin coverage one can
successfully carry out parameter estimations with results,
at least as good as with the phenomenological models
\cite{TheLIGOScientific:2016wfe}. In particular, in this third
RIT Catalog release we included coverage of spins above 0.95 in
magnitude up to mass ratios 2:1. 

New forthcoming simulation produced (for instance targeted to
followup any new detection or catalog expansions) will contribute to
improve the binary parameter coverage, thus reducing the interpolation
error. The next step will be reduce the extrapolation error
at very high spins by adding more simulations with spin magnitudes 
above 0.95. Also the extension of the family of simulations displayed
in Fig. \ref{fig:precpanels} to smaller mass ratios, i.e. $q\leq1/3$. 
Coverage for
low total binary masses (below $20M_\odot$), in turn, would require longer
full numerical simulations or hybridization of the current NR waveforms
with post-Newtonian waveforms \cite{Sadiq:2020hti}.

The next area of development for the numerical relativity waveform
catalogs is the coverage of precessing binaries. Those require
expansion of the parameter space to seven dimensions (assuming
negligible eccentricity), and is being carried out in a hierarchical
approach by first neglecting the effects of the spin of the secondary black
holes, which is a good assumption for small mass ratios. This approach
has proven also successful when applied to all O1/O2 events
(J.Healy et al. in preparation 2020).  It required an homogeneous set of
simulations since the differences in $\ln {\cal L}$ are subtle, hence
the need to expand the current RIT catalog.
In a second stage, a follow up of the first determined spin orientations
can be performed with a two spin search.
Another line of extension of the use of NR waveforms is its use in
searches of GW (in addition to that of parameter estimation) a first
implementation of the nonspinning waveforms 
(using for instance the simulations reported in \cite{Healy:2017mvh}
would produce a prototype of this search analysis, and for seach of
recoil effects~\cite{Lousto:2019lyf}.

The simulations of orbiting black-hole binaries
produce also information about the final remnant of the merger of the two
holes.  Numerous empirical formulas relating the initial parameters
$(q,\vec\chi_1,\vec\chi_2)$ (individual masses and spins) of the
binary to those of the final remnant $(m_f,\vec\chi_f,\vec{V}_f)$ have
been proposed. These include formulas for the final mass, spin, and
recoil velocity
\cite{Barausse:2012qz,Rezzolla:2007rz,Hofmann:2016yih,Jimenez-Forteza:2016oae,Lousto:2009mf,Lousto:2013wta,Hemberger:2013hsa,Healy:2014yta,Zlochower:2015wga},
the computation of the peak frequency of the (2,2) mode
$\Omega_{22}^{peak}$, peak waveform
amplitude $A_{22}^{peak}$
~\cite{Healy:2017mvh,Healy:2018swt} and peak luminosity
\cite{TheLIGOScientific:2016wfe,TheLIGOScientific:2016pea,Healy:2016lce,Keitel:2016krm}. Recently also surrogate methods have been reported
in Refs. \cite{Gerosa:2018qay,Varma:2019csw}.
Those formulas in turn provide further tools to extract information
from the observation of gravitational waves, see for instance our
Fig.~\ref{fig:correlations}. The tables in the Appendix \ref{app:ID}
can be used to further model and test remnant and merger waveform
parameters in terms of those of the binary and as a test of gravity.

%%%%%%%%%%%%%%%%%%%%%%%%%%%%%%%%%%%%%%%%%%%%%

\begin{acknowledgments}

The authors thank Y.Zlochower and M.Campanelli for numerous consultations.
The authors also gratefully acknowledge the National Science Foundation
(NSF) for financial support from Grants No.\ PHY-1607520,
No.\ PHY-1707946, No.\ ACI-1550436, No.\ AST-1516150,
No.\ ACI-1516125, No.\ PHY-1726215.  This work used the Extreme
Science and Engineering Discovery Environment (XSEDE) [allocation
  TG-PHY060027N], which is supported by NSF grant No. ACI-1548562.
Computational resources were also provided by the NewHorizons, BlueSky
Clusters, and Green Prairies at the Rochester Institute of Technology,
which were supported by NSF grants No.\ PHY-0722703, No.\ DMS-0820923,
No.\ AST-1028087, No.\ PHY-1229173, and No.\ PHY-1726215.
\end{acknowledgments}

%%%%%%%%%%%%%%%%%%%%%%%%%%%%%%%%%%%%%%%%

\bibliographystyle{apsrev4-1}
\bibliography{../../Bibtex/references}

%%%%%%%%%%%%%%%%%%%%%%%%%%%%%%%%%%%%%%%%%%%%
% Suppressing appendix
%\end{document}
%%%%%%%%%%%%%%%%%%%%%%%%%%%%%%%%%%%%%%%%%%%%
%\newpage

\appendix*
\section{Tables of initial data and results of the new simulations}\label{app:ID}

\begin{figure*}
\includegraphics[angle=0,width=2.00\columnwidth]{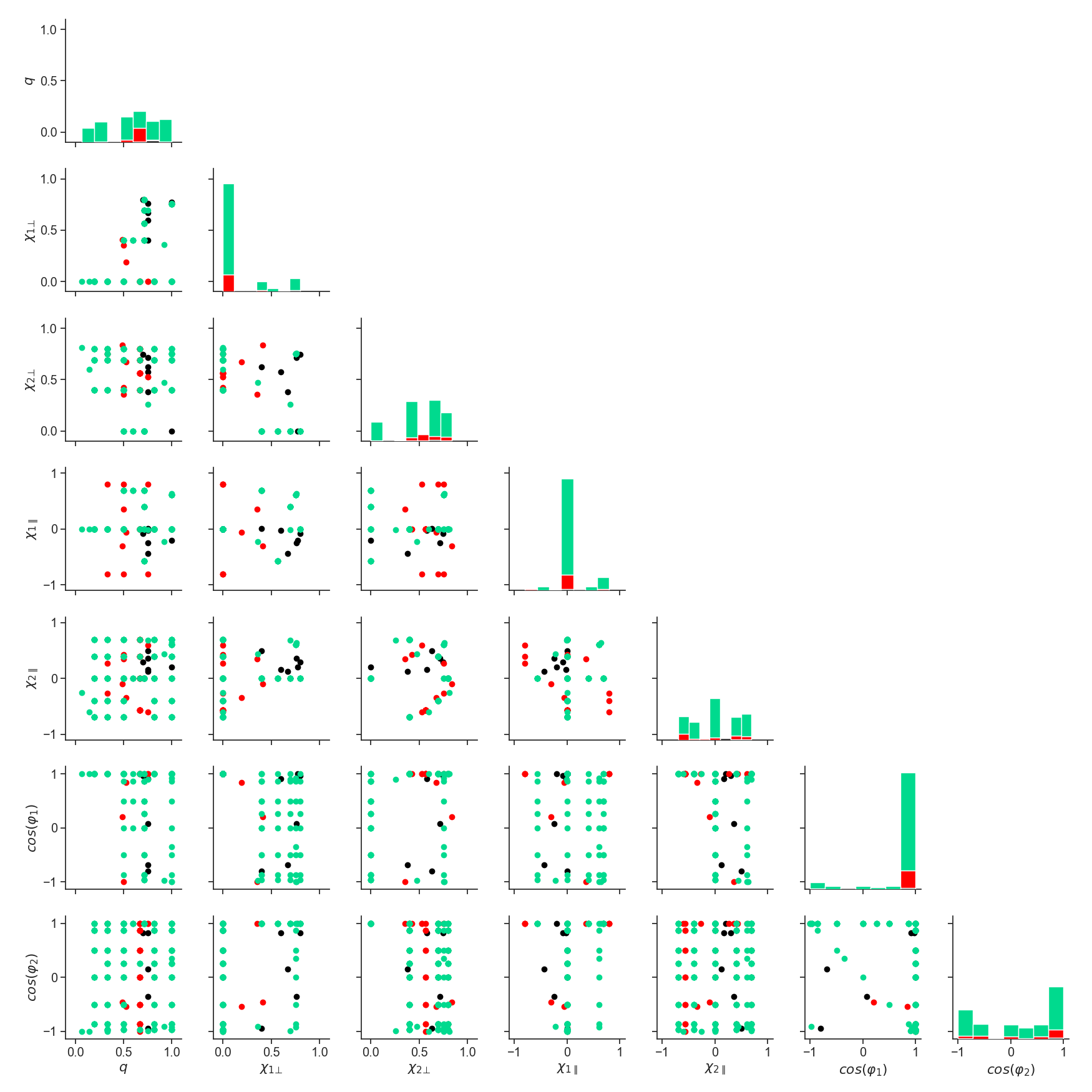}
\caption{Panels showing different combinations of the 7 binary parameters of the precessing parameter space $(q, |\chi_1|, \theta_1, \varphi_1, |\chi_2|, \theta_2, \varphi_2)$ for the 300 precessing simulations (black first release, red second release, blue third release) in this catalog.
\label{fig:precInit}}
\end{figure*}

In this appendix we provide tables with the relevant BBH
configuration details.  In Table \ref{tab:ID}, we provide the
initial data parameters  used to start
the full numerical evolutions. In Tables \ref{tab:IDr} and \ref{tab:IDr_prec},
we provide the binary mass and spin parameters after 
they settle into a more physical value after radiating and absorbing
the spurious gravitation wave content from the initial mathematical 
choice of conformal flatness.  These relaxed values are calculated at  
a fiducial $t=200m$.  A visualization of the 300 binaries'
7 parameter precessing space is displayed in Fig.~\ref{fig:precInit}.

In Table \ref{tab:ecc} we provide the initial orbital frequency
and eccentricity, as well as the number of orbits to merger and the
final eccentricity. The eccentricity is expected to be reduced from its 
initial value by gravitational radiation, at a rate proportional to
$d^{19/12}$ according to \cite{Peters:1964zz}, with $d$, the separation of the
binary (see, for instance, Fig.~6 of Ref.~\cite{Mroue:2010re} or
Fig.~9 in Ref.~\cite{Lousto:2015uwa}).

Finally, In Table \ref{tab:spinerad}, we provide
the values of the energy radiated during the simulation
and the final black hole spin as measured through the (most accurate)
isolated horizon formalism \cite{Dreyer02a}.

%%%%%%%%%%%%%%%%%%%%%%%%%%%%%%%%%%%%%%%%%%%%
%\clearpage

%\clearpage
\begin{longtable*}{lccccccccccccc}
\caption{Initial data parameters for the quasi-circular
configurations with a smaller mass black hole (labeled 1),
and a larger mass spinning black hole (labeled 2). The punctures are located
at $\vec r_1 = (x_1,0,0)$ and $\vec r_2 = (x_2,0,0)$, with individual linear momenta
$P=\pm (P_r, P_t,0)$, spin magnitudes $|S_i|$, puncture mass parameters
$m^p/m$, horizon (Christodoulou) masses $m^H/m$, total ADM mass
$M_{\rm ADM}$, and dimensionless spins $|a/m_H| = |S/m_H^2|$.
The spin directions for the precessing simulations are given in
the catalog and in Tab.~\ref{tab:IDr_prec}.}
\label{tab:ID}\\
\hline
\hline
Run   & $x_1/m$ & $x_2/m$  & $P_r/m$    & $P_t/m$ & $m^p_1/m$ & $m^p_2/m$ & $|S_1/m^2|$ & $|S_2/m^2|$ & $m^H_1/m$ & $m^H_2/m$ & $M_{\rm ADM}/m$ & $|a_1/m_1^H|$ & $|a_2/m_2^H|$\\
\hline
\endfirsthead

\multicolumn{14}{c}
{{\tablename\ \thetable{} -- continued from previous
page}}\\
\hline
Run   & $x_1/m$ & $x_2/m$  & $P_r/m$    & $P_t/m$ & $m^p_1/m$ & $m^p_2/m$ & $|S_1/m^2|$ & $|S_2/m^2|$ & $m^H_1/m$ & $m^H_2/m$ & $M_{\rm ADM}/m$ & $|a_1/m_1^H|$ & $|a_2/m_2^H|$\\
\hline
\endhead
\hline \multicolumn{14}{r}{{Continued on next page}} \\
\hline
\endfoot
\hline\hline
\endlastfoot

RIT:BBH:0275 & -11.59 & 1.66 & -8.72e-05 & 0.03601 & 0.1199 & 0.4593 & 0 & 0.6508 & 0.125 & 0.875 & 0.9966 & 0 & 0.85 \\
RIT:BBH:0282 & -9.38 & 0.625 & -6.03e-05 & 0.02301 & 0.0589 & 0.4929 & 0 & 0.7471 & 0.0625 & 0.9375 & 0.9977 & 0 & 0.85 \\
RIT:BBH:0310 & -6.73 & 4.80 & -5.42e-04 & 0.08356 & 0.2516 & 0.5717 & 0.1389 & 0 & 0.4167 & 0.5833 & 0.9911 & 0.8 & 0 \\
RIT:BBH:0313 & -6.73 & 4.80 & -5.43e-04 & 0.08358 & 0.2516 & 0.5717 & 0.1389 & 0 & 0.4167 & 0.5833 & 0.9911 & 0.8 & 0 \\
RIT:BBH:0315 & -6.73 & 4.80 & -5.43e-04 & 0.08356 & 0.2516 & 0.5717 & 0.1389 & 0 & 0.4167 & 0.5833 & 0.9911 & 0.8 & 0 \\
RIT:BBH:0320 & -6.73 & 4.80 & -5.43e-04 & 0.08359 & 0.2516 & 0.5717 & 0.1389 & 0 & 0.4167 & 0.5833 & 0.9911 & 0.8 & 0 \\
RIT:BBH:0323 & -4.50 & 4.50 & -1.18e-03 & 0.103 & 0.5 & 0.5 & 0.2425 & 0.2425 & 0.5 & 0.5 & 0.988 & 0.97 & 0.97 \\
RIT:BBH:0325 & -6.73 & 4.80 & -5.42e-04 & 0.08356 & 0.2516 & 0.5717 & 0.1389 & 0 & 0.4167 & 0.5833 & 0.9911 & 0.8 & 0 \\
\end{longtable*}

\clearpage
\begin{longtable}{cccccc}
\caption{The masses and spins of the nonprecessing BHBs in Table~\ref{tab:ID} after the
BHs had time to equilibrate ($t/m=200$).  
%Also provided are the difference in the masses,
%$\delta m = (q-1)/(q+1)$, sum of the spins $S/m^2 = (\chi_{2z} + q^2\chi_{1z})/(q+1)^2$, and spin difference, 
%$\Delta/m^2 = (\chi_{2z} - q\chi_{1z})/(1+q)$.
}
\label{tab:IDr}\\
\hline
\hline
Run & $q^r$ & $m^r_1/m$ & $m^r_2/m$ & $\chi^r_{1z}$ & $\chi^r_{2z}$  \\
\hline
\endfirsthead

\multicolumn{6}{c}
{{\tablename\ \thetable{} -- continued from previous page}}\\
\hline
Run & $q^r$ & $m^r_1/m$ & $m^r_2/m$ & $\chi^r_{1z}$ & $\chi^r_{2z}$  \\
\hline
\endhead
\hline \multicolumn{6}{r}{{Continued on next page}} \\
\hline
\endfoot
\hline\hline
\endlastfoot

RIT:BBH:0357 &	$0.6004$ & $0.3750$ & $0.6245$ & $0.5000$ & $-0.8513$ \\
RIT:BBH:0358 &	$0.6001$ & $0.3750$ & $0.6249$ & $0.5000$ & $-0.6502$ \\
RIT:BBH:0365 &	$0.3335$ & $0.2500$ & $0.7496$ & $-0.0000$ & $0.8009$ \\
RIT:BBH:0366 &	$1.0005$ & $0.5000$ & $0.4998$ & $0.0000$ & $-0.8008$ \\
RIT:BBH:0370 &	$0.6672$ & $0.4000$ & $0.5995$ & $-0.5001$ & $-0.8505$ \\
RIT:BBH:0371 &	$0.6672$ & $0.4000$ & $0.5995$ & $0.5000$ & $0.8506$ \\
RIT:BBH:0373 &	$0.0667$ & $0.0625$ & $0.9375$ & $0.0000$ & $0.0000$ \\
RIT:BBH:0374 &	$0.6000$ & $0.3750$ & $0.6250$ & $-0.0000$ & $0.5000$ \\
\end{longtable}

\clearpage
\begin{longtable*}{lcccrrrrrr}
\caption{The mass and spin of the precessing BHBs in Table~\ref{tab:ID} after the
BHs had time to equilibrate ($t/m=200$).}
\label{tab:IDr_prec}\\
\hline
\hline
Run & $q^r$ & $m^r_1/m$ & $m^r_2/m$ & $\chi^r_{1x}$ & $\chi^r_{1y}$ &  $\chi^r_{1z}$ &  $\chi^r_{2x}$ &  $\chi^r_{2y}$ & $\chi^r_{2z}$ \\
\hline
\endfirsthead

\multicolumn{10}{c}
{{\tablename\ \thetable{} -- continued from previous page}}\\
\hline
Run & $q^r$ & $m^r_1/m$ & $m^r_2/m$ & $\chi^r_{1x}$ & $\chi^r_{1y}$ &  $\chi^r_{1z}$ &  $\chi^r_{2x}$ &  $\chi^r_{2y}$ & $\chi^r_{2z}$ \\
\hline
\endhead
\hline \multicolumn{10}{r}{{Continued on next page}} \\
\hline
\endfoot
\hline\hline
\endlastfoot

RIT:BBH:0275 &	$0.1430$ & $0.1250$ & $0.8743$ & - & - & - & -0.6003 & -0.0334 & -0.6028 \\
RIT:BBH:0282 &	$0.0667$ & $0.0625$ & $0.9367$ & - & - & - & -0.8120 & -0.0333 & -0.2538 \\
RIT:BBH:0310 &	$0.7140$ & $0.4165$ & $0.5833$ & 0.3710 & 0.1619 & 0.6908 & - & - & - \\
RIT:BBH:0313 &	$0.7140$ & $0.4165$ & $0.5833$ & 0.0723 & 0.3985 & 0.6906 & - & - & - \\
RIT:BBH:0315 &	$0.7140$ & $0.4165$ & $0.5833$ & 0.2591 & 0.3202 & 0.6866 & - & - & - \\
RIT:BBH:0320 &	$0.7140$ & $0.4165$ & $0.5833$ & -0.1246 & 0.3558 & 0.7063 & - & - & - \\
RIT:BBH:0323 &	$1.0000$ & $0.5002$ & $0.5002$ & -0.6913 & -0.2925 & 0.6166 & 0.6912 & 0.2925 & 0.6166 \\
RIT:BBH:0325 &	$0.7140$ & $0.4165$ & $0.5833$ & -0.3871 & 0.0396 & 0.6997 & - & - & - \\
\end{longtable*}

\clearpage
\begin{longtable}{lcccc}
\caption{Table of the initial orbital frequency $m\omega_i$,
number of orbits to merger, $N$, and the initial and final eccentricities,
$e_i$ and $e_f$ for the spinning cases.}
\label{tab:ecc}\\
\hline
\hline
Run & $m\omega_i$ & $N$ & $e_i$ & $e_f$ \\
\hline
\endfirsthead

\multicolumn{5}{c}
{{\tablename\ \thetable{} -- continued from previous page}}\\
\hline
Run & $m\omega_i$ & $N$ & $e_i$ & $e_f$ \\
\hline
\endhead
\hline \multicolumn{5}{r}{{Continued on next page}} \\
\hline
\endfoot
\hline\hline
\endlastfoot

RIT:BBH:0275 &	$0.0192$ & $18.0$ & $0.0038$ & $0.0037$\\
RIT:BBH:0282 &	$0.0280$ & $14.6$ & $0.0066$ & $0.0076$\\
RIT:BBH:0310 &	$0.0236$ & $10.4$ & $0.0029$ & $0.0016$\\
RIT:BBH:0313 &	$0.0236$ & $10.5$ & $0.0040$ & $0.0014$\\
RIT:BBH:0315 &	$0.0236$ & $10.5$ & $0.0069$ & $0.0016$\\
RIT:BBH:0320 &	$0.0236$ & $10.5$ & $0.0070$ & $0.0014$\\
RIT:BBH:0323 &	$0.0340$ & $7.0$ & $0.0075$ & $0.0027$\\
RIT:BBH:0325 &	$0.0236$ & $10.6$ & $0.0033$ & $0.0015$\\
\end{longtable}

\clearpage
\begin{longtable}{lcr}
\caption{The energy radiated, $\delta \mathcal{M}^{IH} = M_{adm} - M_{rem}^{IH}$,
and final spin, $\chi_{\mathrm{rem}}^{IH}$, as measured using the IH formalism
The error bars are due to variations in the measured mass and spin with time.  
For aligned systems with final spin antialigned to the initial orbital angular 
momentum the minus sign is preserved.}
\label{tab:spinerad}\\
\hline
\hline
Run & $\delta \mathcal{M}^{IH}$ & $\chi_{\mathrm{rem}}^{IH}$ \\
\hline
\hline
\endfirsthead

\multicolumn{3}{c}
{{\tablename\ \thetable{} -- continued from previous page}}\\
\hline
Run & $\delta \mathcal{M}^{IH}$ & $\chi_{\mathrm{rem}}^{IH}$ \\
\hline
\endhead
\hline \multicolumn{3}{r}{{Continued on next page}} \\
\hline
\endfoot
\hline\hline
\endlastfoot

RIT:BBH:0275 &  $0.009755 \pm 0.000021$ & 	$0.466308 \pm 0.000064$ \\
RIT:BBH:0282 &  $0.005172 \pm 0.000045$ & 	$0.716323 \pm 0.000227$ \\
RIT:BBH:0310 &  $0.053655 \pm 0.000000$ & 	$0.738437 \pm 0.000001$ \\
RIT:BBH:0313 &  $0.053591 \pm 0.000000$ & 	$0.738922 \pm 0.000000$ \\
RIT:BBH:0315 &  $0.053710 \pm 0.000000$ & 	$0.738432 \pm 0.000000$ \\
RIT:BBH:0320 &  $0.053412 \pm 0.000000$ & 	$0.739440 \pm 0.000001$ \\
RIT:BBH:0323 &  $0.075869 \pm 0.000002$ & 	$0.852506 \pm 0.000002$ \\
RIT:BBH:0325 &  $0.053473 \pm 0.000000$ & 	$0.738955 \pm 0.000000$ \\
\end{longtable}

%%%%%%%%%%%%%%%%%%%%%%%%%%%%%%%%%%%%%%%%%%%%

\end{document}